# Bifurcation patterns of market regime transition

SERGEY KAMENSHCHIKOV

Moscow State University of M.V.Lomonosov, Faculty of Physics,
Russian Federation, Moscow, Leninskie Gory, Moscow, 119991

IFC Markets Corp., Analytics division, UK, London,
145-157 St John Street, EC1V 4PY

In this paper mechanisms of reversion - momentum transition are considered. Two basic nonlinear mechanisms are highlighted: a slow and fast bifurcation. A slow bifurcation leads to the equilibrium evolution, preceded by stability loss delay of a control parameter. A single order parameter is introduced by Markovian chain diffusion, which plays a role of a precursor. A fast bifurcation is formed by a singular fusion of unstable and stable equilibrium states. The effect of a precatastrophic range compression is observed before the discrete change of a system. A diffusion time scaling is presented as a precursor of the fast bifurcation. The efficiency of both precursors in a currency market was illustrated by simulation of a prototype of a trading system.

## 1. Introduction

Two general classes may be considered among the quantitative trading systems: mean reversion and momentum based systems. The mean reversion model assumes a stationarity of returns and an existence of a high probability attraction range. A price reversion inside this range discloses trading opportunities. Normal Gaussian distribution is considered as a preliminary model of small deviations, which makes traditional statistical techniques easy to apply. Popularity of traditional systems led to a high competition in algorithmic trading industry. As a result, the mean reversion opportunities tend to collapse in short intervals. This tendency forced some of quantitative funds and individual traders to focus on the momentum based models. These models assume market inefficiency, nonstationary behavior of returns and long-term memory existence. In this case, a normal distribution is less effective for description because of "fat tails", considered by Vilfredo Pareto.

A quantitative approach commonly utilizes these models separately for trading mean-reverting portfolios or momentum systems. However, fractal analysis (see Mandelbrot (1968)) shows that an evolving market may exist in two modes simultaneously for several time scales: a fractal random walk may be considered as a superposition of momentum and mean reversion stages. Absence of a stable time scale makes the alternation of mean reversion and momentum modes inevitable. That is why an algorithm type should be switched in time, and corresponding corrections into risk management have to be involved. The purpose of this research is to find out nonlinear fundamental effects of transitions, which help to forecast a switching point.

## 2. Two mechanisms of transition

Models of transition may be classified based on properties of equilibrium surface *P(R)*, which expresses the relation between price(s) and control parameter(s). To build this curve we assume that the market state is defined by a set of control fundamental parameters $R_i, i = \overline{1, N}$ (P/E for stocks, basic rate for currencies, etc). The qualitative state may be defined geometrically as an area of extreme probability density. For the one-dimensional case two states of a system are illustrated in figure 1 and figure 2 as S1 and S2. The state curve, presented in figure 1, is continuous and S1-S2 transition may be denoted as continuous, slow bifurcation. In this case, the system passes quasi-stable transitional states between D and E before a new equilibrium is reached. A slow transition is possible if characteristic period of parameter change is higher than period of system relaxation to quasi-stable transitional state: $\tau_R \succ\succ T$. A gradual diffusion of information in the market is an example of a slow transition.

Another type of transition corresponds to discontinuity, presented in figure 2. The "cusp" D-C-A-B-E was analyzed by Hassler Whitney (1955) in the frame of a catastrophe theory. It is formed by deformation of $P(R)$ when two stable attracting regions S1 and S2 converge and one of them looses stability. Before the curve deformation D, C, A, B corresponds to different parameter values. However, this deformation leads to multiplicity and singularity. According to Whitney, the disruption $D \to E$ appears as a fusion of stable and unstable regimes, marked by ovals. It corresponds to the fast bifurcation because a small deviation of control parameter $R$ immediately leads to discrete "flight" from D-state into E-state or vice versa. This flight is possible when characteristic period of this parameter change is less than the period of system relaxation to quasi-stable transitional state: $\tau_R \prec\prec T$. Here intermediate states cannot exist. Discrete and fast injection of information may lead to this type of transition. For example, it relates to a small unpredictable deviation of fundamental indicator from consensus of analysts.

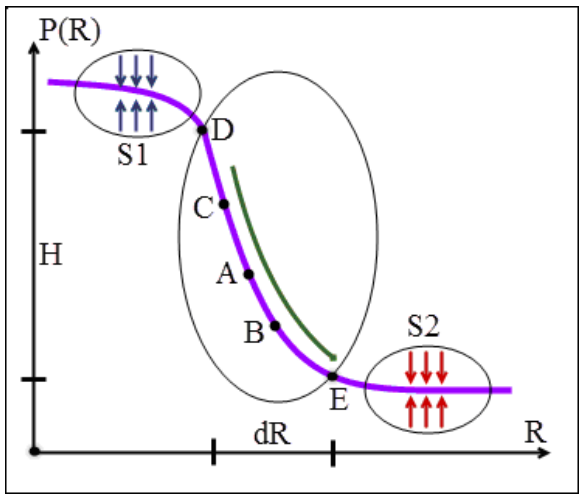 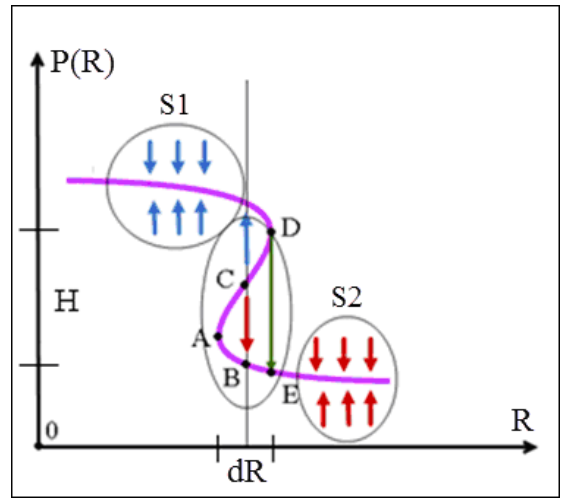

Figire 1. Slow transition                Figure 2. Fast transition

Below we consider two fundamental effects, which help to forecast slow and fast transitions separately.

## 3. Slow transition

The analytical basis of a slow, quasi-stable transition was developed by Neishtadt (1987). He assumed that control parameter changed slowly in relation to the system characteristic time: $\tau_R \succ\succ T$. Information is gradually diffusing and $P(R)$ mapping curve is continuous. Let us consider a nonlinear system (2). We may introduce a new variable of price and the dynamics law:

$$P(t) = P_0 + \xi(t). \qquad (1)$$

$$\dot{P} = f(P,R) \quad \dot{R} = \varepsilon \quad \varepsilon \prec\prec 1 \quad \delta R \prec\prec R \qquad (2)$$

The equation may be modified using decomposition:

$$\dot{\xi} = A(R)\xi + O(|\xi^2|) + \varepsilon h(R) \quad h \prec\prec 1 \quad \xi = P - P_0 \qquad (3)$$

Here $(P_0, R_0)$ corresponds to some equilibrium state. A critical $R^*$ corresponds to the positive imaginary part $A(R)$, while for $R \prec R^*$ $A(R)$ has only negative imaginary solutions. Neishtadt (1987) showed that a phase point of $P(T_0)$ corresponding to $R^* - dR$ should be attracted to the instability regime eventually. Time needed for instability increase is defined by the adiabatic parameter.

$$T_R \approx \dot{R} \quad \xi(T_R)/\xi(T_0) > 1 \tag{4}$$

This relation means that a slow bifurcation is preceded by a control parameter increase: stability loss delay (SLD-effect).

## 3.1. Transport model

The choice of a control parameter is a contentious issue itself and depends on the selected transport model. In case of strong mean-reverting stable and quasi stable states a Markovian short-term memory model may be successfully applied. Indeed, in case of strong mean reversion autocorrelations and trends quickly disappear in vicinity of turning points. This condition can be presented as: $T_R \prec\prec T_0$. Here $T_R$ is autocorrelations decay period and $T_0$ is a characteristic period of a system evolution. If we consider a one-dimensional case, a transitional probability for short memory random walks formally satisfies Chapman - Kolmogorov relation:

$$W(p_3,t_3 \mid p_1,t_1) = \int dp_2 W(p_3,t_3 \mid p_2,t_2) W(p_2,t_2 \mid p_1,t_1) \tag{5}$$

Here $W(p,t \mid p_0,t_0)$ is a conditional probability density. Fokker – Planck – Kolmogorov (FPK) equation is based on two basic assumptions: transitional probability and covariations do not depend on initial time point; and final probability does not depend on the initial coordinate (short memory conditions): $W(p',t' \mid p,t) = W(p',p,t'-t)$, $W(p',p,t) = W(p',t)$.

FPK equation than may be presented in the following way:

$$\frac{\partial W(p,t)}{\partial t} = \frac{1}{2}\frac{\partial}{\partial p}\left(D(p)\frac{\partial W(p,t)}{\partial t}\right) \quad D(p) = \lim_{\Delta t \to 0}\left(\frac{\langle\langle_\Delta p^2\rangle\rangle}{\Delta t}\right) \tag{6}$$

Here double brackets designate an averaging in relation to the initial price-coordinate:

$$\langle\langle_\Delta p^2\rangle\rangle = \int (p-p_0)^2 W(p,p_0,_\Delta t) dp_0 \tag{7}$$

This model is applicable for a slow transition and a continuous state curve when correspondence between price, control parameter and time is ambiguous: diffusion has an implicit hidden time parameter - $D(p) = D(p(R(t)))$. Diffusion may be expressed by the absorbed energy as well:

$$\varepsilon(p) = \left(\frac{_\Delta p(p)}{_\Delta t_{min}}\right)^2 \quad D(p) = \frac{\langle\langle_\Delta p^2\rangle\rangle}{_\Delta t_{min}} = _\Delta t_{min}\langle\langle\varepsilon(p)\rangle\rangle \tag{8}$$

In case of Markovian short memory approach an averaged energy $\langle\langle\varepsilon(p)\rangle\rangle$ may be used as a single control parameter, which defines dynamical properties according to the equation (6). The change of control parameter is expressed by the derivative:

$$S = sign\left(\frac{d\langle\langle\varepsilon\rangle\rangle}{dt}\right) = sign\left(\frac{dD}{dt}\right) \tag{9}$$

The core of this hypothesis is that a mean reversion to momentum slow transition is preceded by positive S precursor, which expresses Neishtadt attraction to an instable regime. Positive S corresponds to growing fluctuations (averaged energy and diffusion). It means growth of volatility due to gradual diffusion of information for market applications.

## 3.2. Market patterns

There are several ways of SLD interpretation in quantitative trading. Below one of trivial descriptions is suggested. It allows introducing of a trading prototype, which may be used for the development of a "full-version" adaptive algorithm.
The mechanism of stability loss delay may be considered in a frame of suggested transport model when four qualitative assumptions are satisfied:

- Short-term memory of mean reversion dynamics;
- Slow growth of a control parameter;
- Delay: the system has not bifurcated but has already lost stability;
- Transition finally leads to a new equilibrium.

An existence of a principle time scale $T_0$ is assumed to simplify illustration of SLD. However a working trading algorithm certainly will require an adaptation scheme. Our analysis is based on uniform time series of the close prices. A principle time scale $T_0$ corresponds to $N$ number of points. Below a trivial formalization of qualitative requirements is suggested:

1. Auto correlation function decay: $ACF(T_0, T_0) \prec\prec ACF^*$. Here $ACF(T_0, T_0)$ shows a correlation factor between two equal time series $(p_{i-N},\ldots,p_i)$, $(p_{i-2N},\ldots,p_{i-N})$ of equal length $N$ - one of them shifted by $N$ points backwards. $ACF^*$ is some relatively small quantity ($0 \prec ACF^* \prec 1$), which is a parameter of an algorithm;

2. A slow growth of a control parameter: $0 \prec (D_i - D_{i-1}) \prec \theta \cdot D_i$. Here $\theta$ is some small factor ($0 \prec \theta \prec 1$) – parameter of the algorithm. The diffusion is estimated according to the following relation (see relation 8):

$$D_i = \frac{1}{N^2} \sum_{j=i-N}^{i} (p_i - p_0)^2 \qquad (10)$$

3. A proximity to the mean reversion state is described by a natural requirement $|p_i - \mu_i| \prec F \cdot \sigma_i$ with free $F$ parameter. Here a standard deviation $\sigma_i$ and average $\mu_i$ are estimated according to the trivial relations:

$$\mu_i = \frac{1}{N} \sum_{j=i-N}^{i} p_j \quad \sigma_i = \sqrt{\frac{1}{N-1} \sum_{j=i-N}^{i} (p_j - \mu_i)^2} \qquad (11)$$

4. A new equilibrium state is described by two conditions. First one corresponds to a simple reversion - $p_i \succ \mu_i$ (after bearish transition), $p_i \prec \mu_i$ (after bullish transition). Another one expresses a decrease of fluctuations and diffusion compression: $(D_i - D_{i-1}) \prec 0$.

A trade position is opened when 1-3 requirements are satisfied. The direction of probable transition is defined according to the recent moving averages: $p_i \succ \mu_i$ (bullish transition), $p_i \prec \mu_i$ (bearish transition). The position is closed if the fourth requirement is met. This trivial scheme assumes an instant execution according to the close price (market order) and perfect liquidity. Certainly this simplification should be avoided when constructing a "full-version" algorithm. Below MatLab results of simulation for British pound are represented: cumulated yield vice number of trades. Four digits quotes are provided by Dukascopy Bank with average transaction costs: bid/ask spread=1.8 pips, overnight swap long=0.38 pips, overnight swap short=0.11 pips. "Four hours" timeframe is selected – "Close" price is written down with four hours time resolution from July 2010 till November 2015. Overall historical volume therefore includes 7950 points. There is no risk management in a prototype scheme: the equal trading volume of $100,000 is preserved in each trade. Four free parameters of algorithm are optimized up to trade №110 in relation to maximal cumulated yield: N=5, F=1.4, $\theta = 0.1$ and $ACF^*$=0.7. A linear regression of Yield (Trades) curve gives a determination factor of $R^2 = 92\%$. A positive statistical significance of a determination factor is observed according to Fisher test.

The relation of a maximum cumulated yield 16% (4% per year), and a maximum drawdown of 2% is optimistic - Ca=8. However, there are three platforms (8 months duration each), revealing low efficiency stages of the algorithm. They are emphasized by nine degree polynomial approximation curve. In spite of the fact that out-of-sample test is included into the simulation (green rectangular), it does not provide a long-term stability in future. According to the fractal nature of the markets (see Mandelbrot (1968)), there is no a long-term time scale in the liquid markets. That is why our "optimal" scale (N=5) has a limited efficiency – the prototype model should be supplemented by adaptive elements in the practical solutions.

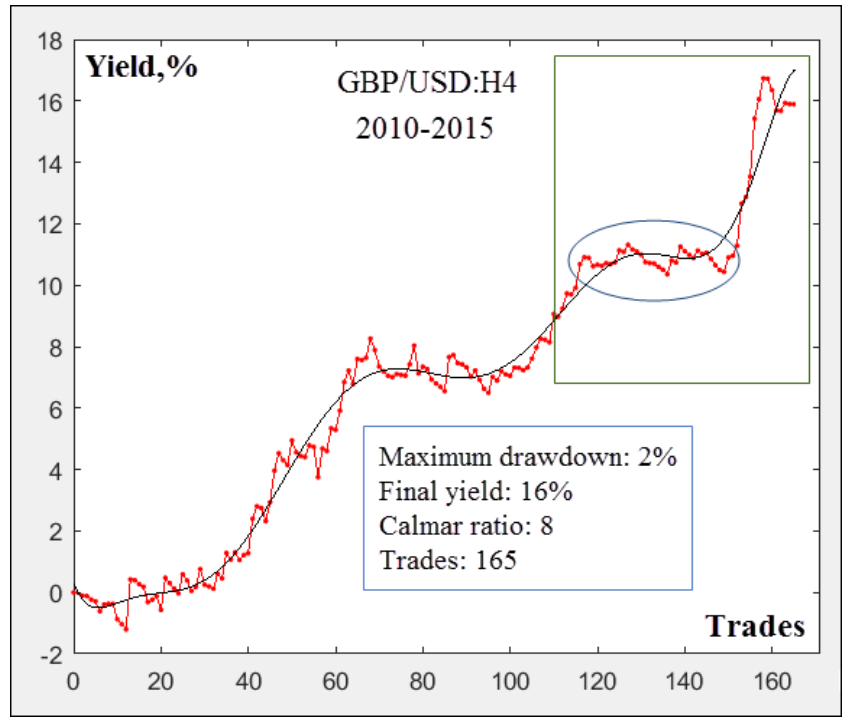

Figire 3. MatLab simulation: cumulated yield (red) and polynomial approximation (black).
Green rectangular marks out-of-sample simulation: 53 trades.

**4. Fast transition**

Another type of transition is presented by a singular disruption $D \to E$ in figure 2. Let us consider an irreversible mechanism, which may lead to this singularity. Poincare showed that any dynamic state can be modeled as a combination of stable points (focuses), periodic/quasi-periodic fluctuations and unstable trajectories (separatrixes).

Andronov (1937) and Leontovich proved that there are only five types of fast irreversible bifurcations:

- **Unstable quasi periodic motion $\Rightarrow$ stable focus;**
- **Stable focus $\Rightarrow$ unstable focus $\Rightarrow$ stable quasi periodic motion;**
- Unstable focus $\Rightarrow$ stable focus;
- Stable focus $\Rightarrow$ unstable focus;
- Stable quasi periodic motion $\Rightarrow$ unstable quasi periodic motion.

First and second chains are explained by Poincare–Bendixson (1901) theorem which states that irreversible fast transitions of attractors must be preceded by the attraction range compression. This theorem allows interpreting the first and second chain in the following way: an attracting range of quasi-periodic fluctuations is compressed into the small attracting spot – focus (the first chain above). This attracting area is so small that it exists only till a random fluctuation breaks the system out of this range into the next stable regime (the second chain). This effect of precatastrophic compression (PC - effect) is observed in mechanical systems such as turbulent

flows – small-scale intermittency before a large-scale disturbance. It is revealed in the market systems as well through the volatility clustering (see Lux (2000)).

### 4.1. Transport model

A quantitative model of this effect has to include a multiplicity between a control parameter, time and price: $t_0 \leftrightarrow R_0 \to (p_1, p_0,...)$. It means that all quantities of this model have at least two independent variables: $(p,t)$ or $(p,R)$; and initial stationarity requirement of Markov chain is violated: $W(p',t'|p,t) \neq W(p',p,t')$. Non-stationary FPK model, however, preserves some initial properties (see Kamenshchikov (2014)) of traditional Markovian scheme:

$$D(p,t) \approx \left( \frac{\langle\langle_\Delta p^2 \rangle\rangle}{_\Delta t_{min}} \right) \quad _\Delta t_{min} \succ T \tag{12}$$

$$\frac{\partial W(p,t)}{\partial t} \approx \frac{1}{2} \frac{\partial}{\partial p}\left( D(p,t) \frac{\partial W(p,t)}{\partial t} \right) \tag{13}$$

An asymptotical relation for the large scales $(t-t_0) \succ\succ T$ may be presented in the following way:

$$\langle\langle_\Delta p^2 \rangle\rangle = D(p,t)(t-t_0) =_\Delta t \langle\langle \varepsilon(p,t) \rangle\rangle \tag{14}$$

This model assumes non-stationary dynamic properties, which may be caused by external influences such as new fundamentals in the markets, a boundary variation (currency corridor change), a volatility injection, etc. Fractal Brownian motion is a particular case of such dynamics - let us compare (14) with an expectation of fractal Brownian motion Mandelbrot (1968):

$$E\left([B_H(t+T) - B_H(t)]^2\right) = V_H \cdot T^{2H} \tag{15}$$

Here $T = t - t_0$ is a time lag and $H$ is Hurst exponent, which defines the regime type. Cases of $H \succ 0.5$ and $H \prec 0.5$ correspond to the momentum and mean reverting regimes correspondingly. $H = 0.5$ condition corresponds to the stationary case ($D(p,t) = const$) of Wiener process. In general case a transport factor still may be used as a principal control parameter. Than a precatastophic compression (PC - effect) may be described in terms of the transport properties of diffusion or stochastic energy. A compression of an attraction area $\langle\langle_\Delta p^2 \rangle\rangle \to 0$ can be presented asymptotically (16): $D(p,t)(t-t_0) \to 0$.

$$\lim_{t \to \infty} D(p,t) \approx \frac{const}{(t-t_0)^{(F+1)}} \quad F \succ 0 \tag{16}$$

This condition (16) is valid only for $_\Delta D(p,t) \prec 0$ mean reverting regimes of quasi-stable fluctuations ($H \prec 0.5$). Our hypothesis is based on the following: the mean reversion to momentum fast transition is preceded by a positive F precursor, corresponding to a precatastophic compression of the fluctuations. It leads to the rapid decrease of the volatility before the fundamental systematic shifts for the market applications.

### 4.2. Market patterns

There are several ways of PC-effect interpretation in the quantitative trading. Below one of the trivial descriptions is suggested. It allows introducing a trading prototype which may be used for the development of a "full-version" adaptive algorithm.
A pattern of precatastophic stabilization may be considered in the frame of a suggested transport model when four qualitative assumptions are satisfied:
- Short-term memory of mean reversion dynamics and Markovian description;
- Preliminary compression of diffusion before bifurcation;

- Breakdown/bifurcation of mean-reversion state;
- Formation of a new equilibrium/mean reversion state.

An existence of a principle time scale $T_0$ is assumed to simplify an illustration of PC - effect. However, a working trading algorithm certainly requires adaptation scheme. Our analysis is based on the uniform time series of close prices. A principle time scale $T_0$ corresponds to $N$ number of points. Below a trivial formalization of qualitative requirements is suggested:

1. The auto correlation function decay: $ACF(T_0,T_0) \prec\prec ACF^*$. Here $ACF(T_0,T_0)$ presents a correlation factor between two equal time series $(p_{i-N},…,p_i)$, $(p_{i-2N},…,p_{i-N})$ of equal length $N$ - one of them shifted by $N$ points backwards. $ACF^*$ is some small factor ($0 \prec ACF^* \prec 1$) – parameter of the algorithm;
2. The compression of the diffusion is estimated according to the linear regression (11) when a negative bias $\chi \prec 0$ is reached. This regression describes locally the power law of (13). Counter $j$ is varied in the range of N points, while $j_0$ – between j-2N and j-N:

$$\langle\langle \Delta p^2 \rangle\rangle = \sum_{k=j_0}^{j}(p_j - p_k)^2 = f(j,j_0) \qquad (17)$$

$$f(j,j_0) \cong \chi j + const \quad \chi \prec 0 \qquad (18)$$

The compression is verified in $j$ iteration, if $j+1$ iteration corresponds to the breakdown itself;
3. The breakdown of the mean reversion state is described by a natural requirement $|p_i - \mu_i| \succ F\sigma_i$ with free parameter $F$. Here a standard deviation $\sigma_i$ and an average $\mu_i$ are estimated according to the trivial relations:

$$\mu_i = \frac{1}{N}\sum_{j=i-N}^{i} p_j \quad \sigma_i = \sqrt{\frac{1}{N-1}\sum_{j=i-N}^{i}(p_j - \mu_i)^2} \qquad (19)$$

4. A new equilibrium state is described by three conditions. The first condition corresponds to the reversion - $p_i \succ \mu_i$ (after bearish transition), $p_i \prec \mu_i$ (after bullish transition). The second condition expresses a decrease of fluctuations and diffusion compression: $\chi \prec 0$. Finally, eligibility of a transport model has to be verified by a short memory condition: $ACF(T_0,T_0) \prec\prec ACF^*$.

A trade position is opened when 1-3 requirements are satisfied. The direction of a probable transition is defined according to the recent moving averages: $p_i \succ \mu_i$ (bullish transition), $p_i \prec \mu_i$ (bearish transition). The position is closed if the fourth requirement is met. This trivial scheme assumes an instant execution according to the close price (market order) and perfect liquidity. Certainly, this simplification should be avoided when constructing a "full-version" algorithm.

Below **MatLab** results of the simulation for British pound are represented: a cumulated yield vice number of trades. Four digits quotes are provided by Dukascopy Bank with average transaction costs: bid/ask spread=1.8 pips, overnight swap long=0.38 pips, overnight swap short=0.11 pips. "Four hours" timeframe is selected – "Close" price is written down with four hours time resolution from July 2010 till November 2015. Overall historical volume includes 7950 points. There is no risk management in a prototype scheme: the equal trading volume of $100,000 is preserved in each trade. Four free parameters of the algorithm are optimized up to trade №252 in the relation to the maximal cumulated yield: N=5, F=1.3 and $ACF^*$=0.7. A linear regression of Yield (Trades) curve gives a determination factor of $R^2$=83. A positive statistical significance of determination factor is observed according to Fisher test.
The relation between a maximum cumulated yield 17% (4% per year), and a maximum drawdown of 5% is optimistic - Ca=3. However, there are two platforms (9-10 months duration each), revealing the low efficiency stages of the algorithm. They are emphasized by nine degree polynomial approximation curve. In spite of the fact, that out-of-sample test is included into the simulation (green rectangular), it does not provide a long-term stability in future. According to the fractal nature of markets (see Mandelbrot (1968)) there is no long-term time scale in the liquid markets. That is why our "optimal" scale has limited efficiency – the prototype model should be supplemented by the adaptive elements in the practical solutions.

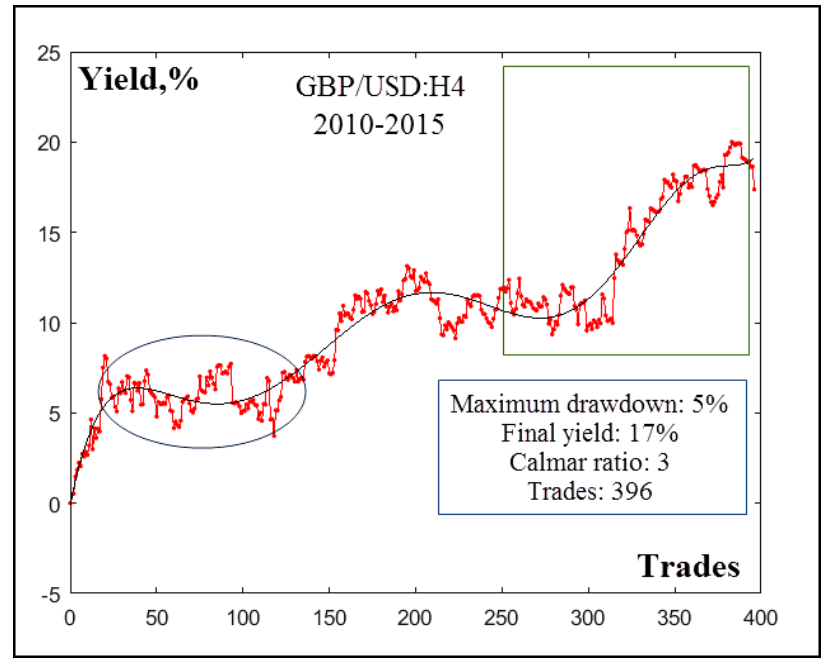

Figure 4. MatLab simulation: cumulated yield (red) and polynomial approximation (black). Green rectangular marks out-of-sample simulation: 396 trades.

## 5. Conclusions

A fractal hypothesis states that an evolving market exists in several scales simultaneously - absence of a stable time scale makes an alternation of mean reversion/momentum modes inevitable. The purpose of this research is to find mechanisms of nonlinear transitions, which could help to detect a switching point. Two patterns of nonlinear transitions are outlined: a slow and fast transition. The slow transition is possible when a parameter change is slower than a system relaxation. A gradual shift of a control parameter is followed by a delay before a disruption of a reversion. A fast transition arises through the singular fusion of unstable and stable equilibriums. It is shown that such transition is preceded by passing through the area of the extremely small fluctuations. Time series interpretations of both transitional patterns are introduced. Markovian model is used to describe transport properties through a single control parameter of the diffusion. The mechanism of a slow transition may be considered when the following assumptions are satisfied: a short-term memory, a slow growth of diffusion, a delay between the fluctuations growth and transition. Markovian approach is extended into a non-stationary area of the fast transitions. Connection between fractal Hurst factor, fluctuations of energy and a transport factor are outlined. The pattern may be considered when the following conditions are satisfied: a short-term memory, a preliminary compression of the diffusion, a breakdown/bifurcation of mean-reversion state.

Both patterns are tested by a simulation of British pound back test trading. There is no risk management and parametric adaptation into this prototype algorithm. Four free parameters and more than hundred trades allow achieving a statistical significance. Out-of-sample optimization is used to demonstrate a local stability of the algorithms. Calmar ratios of slow transition and fast transition algorithms are correspondingly Ca=8 and Ca=3, which proves a local prediction efficiency. We can notice that a slow transition is more accurately detected in the frame of the suggested transport model in relation to the fast bifurcations. Both algorithms should be complemented by the adaptive/self-learning modules to preserve stability in future. The mechanism of the fast transition allows looking at the fractal analysis from the novel fundamental point of view. This interpretation may help traders to improve the statistical techniques and predictive power of applications.